\documentclass[aps,twocolumn,superscriptaddress]{revtex4-2}

\usepackage{graphicx}
\usepackage{empheq}
\usepackage{amsmath}
\usepackage{amsfonts}
\usepackage{amssymb}
\usepackage{amsthm}
\usepackage{dsfont}
\usepackage{mathtools}
\usepackage{float}
\usepackage{color}
\usepackage{tabularx}

\newcommand{\beq}{\begin{equation}}
\newcommand{\eeq}{\end{equation}}

\begin{document}

\author{Christian Arenz} 
\affiliation{Frick Laboratory, Princeton University, Princeton NJ 08544, United States}

\author{Herschel Rabitz} 
\affiliation{Frick Laboratory, Princeton University, Princeton NJ 08544, United States}

\title{Drawing together control landscape and tomography principles}

\date{\today}

\begin{abstract}
The ability to control quantum systems using shaped fields as well as to infer the states of such controlled systems from measurement data are key tasks in the design and operation of quantum devices. Here we associate the success of performing both tasks to the structure of the underlying control landscape. We relate the ability to control and reconstruct the full state of the system to the absence of singular controls, and show that for sufficiently long evolution times singular controls rarely occur. Based on these findings, we describe a learning algorithm for finding optimal controls that makes use of measurement data obtained from partially accessing the system. Open challenges stemming from the concentration of measure phenomenon in high dimensional systems are discussed. 
\end{abstract}

\maketitle

\section{Introduction}

 While quantum control and quantum tomography each have long and distinct research histories, both areas are deeply intertwined. Indeed, manipulating the dynamics of a quantum system in a desired way and reconstructing quantum states through expectation value measurements can be considered as ``two sides of the same coin" \cite{deutsch2010quantum}. In this work we explore the underlying principles drawing both research areas together. By identifying their common foundations, we believe that methods established within each field can be tied together in order to establish new avenues for controlling and reading out quantum systems. 
 
With the advent of laser technologies and pulse shaping techniques, the dream of controlling complex quantum systems was born. Today, preparing a quantum system in a desired state or implementing unitary gates can be achieved through properly tailored control fields \cite{brif2010control, glaser2015training, ball2020software}. Finding the corresponding field shapes through iterative optimization procedures can be accomplished surprisingly effectively, both in numerical simulations \cite{moore2011exploring, moore2011search, floether2012robust, arenz2014control} and laboratory learning control experiments \cite{roslund2009gradient, roslund2009experimental, sun2015experimental, dolde2014high, yang2019silicon, waldherr2014quantum, lu2017enhancing}. The ease of finding such optimal controls is determined by the topology of the underlying control landscape \cite{rabitz2004quantum, chakrabarti2007quantum} defined by a cost functional $J$, often taken to be e.g. the overlap with a target state or the distance from a target unitary transformation. Investigating the properties of the control landscape has therefore attracted much attention \cite{rabitz2004quantum, chakrabarti2007quantum, rabitz2006optimal,  ho2006effective, hsieh2008optimal, pechen2011there, rabitz2012comment, pechen2011there, wu2008control, ho2012critical, wu2012singularities, riviello2014searching, de2013closer, russell2017control, kosut2019quantum}. These research efforts recently culminated in a theorem stating that with sufficient control resources, the control landscape should be free from traps for almost all controlled quantum systems \cite{russell2017control}. However, the precise meaning of ``sufficient" is application dependent and remains an open challenge to systematically assess. It is  known that these issues are related to the ability to steer the dynamics of the system in all directions (in the corresponding tangent space) \cite{brockett1989least,joe2016choosing}. In particular, the appearance of so called \emph{singular controls} \cite{wu2012singularities, riviello2014searching, de2013closer, pechen2011there, rabitz2012comment, pechen2011there}, which hinder the ability to steer the dynamics in all directions, play an essential role. Here, we show that such singular controls manifest itself as measurement data that does not carry sufficient information to  (uniquely) reconstruct generic quantum states, thereby providing a link between quantum control and tomography.

Quantum state tomography aims to reconstruct the quantum state of a system by measuring a set of observables \cite{nielsen2002quantum, paris2004quantum}. This can always be accomplished when the set of observables is \emph{informationally complete} \cite{busch1991informationally}. While this field of research coexists alongside the field of quantum control, both areas are related since creating the required observables can be translated into a control problem. That is, observables that are not directly accessible in the laboratory are typically created by rotating accessible observables into the desired ones by control fields or gates sequences \cite{deutsch2010quantum, steffen2006measurement, cai2013large, bradley2019ten, kokail2019self,goerz2014optimal, ferrie2014self}.  Alternatively, one can infer the state of such driven systems directly from the time traces of accessible observables \cite{liu2019pulsed, silberfarb2005quantum, deutsch2010quantum, merkel2010random, cook2014single, shojaee2018optimal, yang2020complete}. In this case, generic state reconstruction is possible when the obtained data is informationally complete. 

Instead of deterministically creating observables, a set of observables randomly created through Haar random unitary transformations can also be used, as this almost always guarantees information completeness 
    \cite{candes2011probabilistic, merkel2010random, gross2010quantum, ohliger2013efficient}. Since a random control field can create a Haar random unitary evolution \cite{banchi2017driven}, applying a random field and measuring the time trace of a \emph{single} observable almost always yields informationally complete measurement data \cite{yang2020complete}. This eliminates the need to optimize control pulses for rotating observables, as almost all control fields allow for complete quantum state tomography. In this work we will connect this observation to the absence of singular controls.

 The remainder of this work is organized as follows. We begin in Sec. \ref{sec:lcandsc} and \ref{sec:standic} by introducing the mathematical framework that allows for connecting the notions of singular controls and information completeness, and show in Sec. \ref{sec:scandic} that control fields yielding informationally complete measurement data cannot be singular. We then use randomness to conclude that both the absence of singular controls as well as information completeness are generic properties in a measure theoretic sense. Based on these observations, we illustrate in Sec. \ref{sec:learningcontrol} how drawing together quantum control and quantum tomography allows for identifying optimal controls when system access is limited. Finally, we show in Sec. \ref{sec:errorsandlandscape} that in high dimensions, the concentration of measure phenomenon \cite{ledoux2001concentration, mcclean2018barren} leads to increased sensitivity to measurement errors for randomized state tomography, and also to a flattening of the quantum control landscape for a commonly employed cost functional.

\section{Quantum control and state tomography}
 
%\emph{Control landscapes and singular controls} -- 
We begin by considering a quantum system with a $d$ dimensional Hilbert space driven by a classical control field $f(t)$. The evolution of the system is governed by the Schr\"{o}dinger equation $\dot{U}_{t}=-iH(t)U_{t}$ for the time evolution operator $U_{t}$,  where we work in units of $\hbar=1$. We assume that the time dependent Hamiltonian $H(t)$ is of the form \cite{d2007introduction}, 
\begin{align}
\label{eq:Hamiltonian}
H(t)=H_{0}+f(t)H_{c},
\end{align} 
where we refer to $H_{0}$ and $H_{c}$ as the \emph{drift} and the \emph{control} Hamiltonian, respectively. While we focus here on a single control field, the following consideration can be generalized in a straightforward manner. We assume that the drift and the control Hamiltonian are traceless such that the solution of the Schr\"odinger equation $U_{T}$ at time $T$, which we refer to as the \emph{endpoint map}, is an element of the special unitary group $\text{SU}(d)$. The quantum control system given by \eqref{eq:Hamiltonian} is said to be fully controllable if every unitary transformation $V\in \text{SU}(d)$ can be created as a solution to the Schr\"{o}dinger equation. For unconstrained control fields this is equivalent to saying that the \emph{dynamical Lie-algebra} $\mathfrak{L}=\text{Lie}(iH_{0},iH_{c})$ created by nested commutators of the drift and the control Hamiltonian and their real linear combinations span the full space, i.e., the special unitary algebra denoted by $\mathfrak{su}(d)$ \cite{d2007introduction}.

\subsection{Control landscapes and singular controls}\label{sec:lcandsc}
A quantum control problem can be formulated as an optimization problem of a cost functional $J[f(t)]$ over the control field $f(t)$. The goal is to find a function that appropriately either maximizes or minimizes $J$ for a fixed evolution time $T$. The ability to solve the optimization problem relies on the structure of the control landscape defined by the cost functional $J$ \cite{rabitz2004quantum, chakrabarti2007quantum}. In particular, the set of \emph{dynamic} critical points at which $\frac{\delta J[f(t)]}{\delta f(t)}=0$, consisting of local and global optima, plays an essential role in characterizing the topology of the control landscape. A highly favorable property of $J$ would be the absence or rareness of traps given by local optima such that, for instance, gradient type algorithms would be effective in finding the global optimum. Since the cost functional is given by $J=J(U_{T}[f(t)])$, the functional derivative of $J$ with respect to the control field takes the form of a  concatenation, formally written as  $\frac{\delta J}{\delta f(t)}=\frac{\partial J}{\partial U_{T}}\circ\frac{\delta U_{T}}{\delta f(t)}$. The first part captures the landscape properties of $J$ as a function of the endpoint map $U_{T}$, which is referred to as the \emph{kinematic} control landscape. It is well known that the kinematic control landscape of typical cost functionals used for state preparation, gate synthesis, and observable control generally consists of global optima and saddles \cite{brockett1989least, chakrabarti2007quantum, rabitz2006optimal, hsieh2008optimal, joe2016choosing}. If the variation of the endpoint map with respect to the control field $\frac{\delta U_{T}}{\delta f(t)}$ is assumed to be full rank, the dynamic and the kinematic critical points coincide, so that the topology of the control landscape is fully characterized by the critical point structure of the kinematic landscape. This is referred to as \emph{local surjectivity}. However, while local surjectivity is commonly assumed in the first place to conclude that the landscape of typical cost functionals is trap free \cite{chakrabarti2007quantum, hsieh2008optimal}, it is an open problem to determine for what type of controls and systems the assumption holds. In fact, a few examples of controlled systems are known for which local surjectivity fails \cite{wu2012singularities, de2013closer}, as these examples exhibit \emph{singular controls}. Since the variation of the endpoint map with respect to the control field can be written as $\frac{\delta U_{T}}{\delta f(t)}=-iU_{T}U_{t}^{\dagger}H_{c}U_{t}$, a singular control is characterized by the existence of a $v\in i\mathfrak{su}(d)$ for which \cite{ho2006effective, ho2012critical}, 
\begin{align}
\label{eq:singularcritical}
\langle v,U_{t}^{\dagger}H_{c}U_{t}\rangle=0,~~~~\forall t\in[0,T], 
\end{align}
where $\langle A,B\rangle=\text{Tr}\{A^{\dagger}B\}$ denotes the Hilbert-Schmidt inner product. As such, a singular control does not allow for varying the endpoint map $U_{T}$ in all directions associated with the tangent space at $U_{T}$ \cite{wu2012singularities, schulte2010gradient}.

\subsection{State tomography and information completeness}\label{sec:standic}
Consider a $d$ dimensional quantum system in an unknown state $\rho$ whose evolution is governed by the Hamiltonian in Eq. \eqref{eq:Hamiltonian}. The time trace $\langle M\rangle_{t}$ of an observable $M$ is given by 
\begin{align}
\label{eq:observable}
\langle M\rangle_{t}=\langle \rho, U_{t}^{\dagger}MU_{t}\rangle,
\end{align}
where we assume without loss of generality that $M$ is traceless. Using the generalized Bloch vector representation \cite{kimura2003bloch, byrd2003characterization}, the initial state can we written as $\rho=\frac{\mathds{1}}{d}+\sum_{m=1}^{d^{2}-1}\rho_{m}B_{m}$. The coefficients $\rho_{m}=\langle \rho,B_{m}\rangle$ are collected in the Bloch vector $\bold{x}_{\rho}=(\rho_{1},\cdots,\rho_{d^{2}-1})$ with $\{B_{m}\}$ being a complete and orthonormal operator basis for $i\mathfrak{su}(d)$. We assume that at $d^{2}-1$ times $t_{n}$ summarized in the set $\mathcal T=\{t_{n}\}$ we obtain $d^{2}-1$ expectation values, which we collect in the vector $\bold{y}=(\langle M\rangle_{t_{1}},\cdots,\langle M\rangle_{t_{d^{2}-1}})$, and to which we refer to as the \emph{measurement record}. The measurement record is determined by the set of equations $\bold{y}=\mathcal M[f(t)]\bold{x}_{\rho}$, indicating here the explicit dependence of the control field $f(t)$ on the matrix $\mathcal M\in \mathbb R^{(d^{2}-1)\times (d^{2}-1)}$ with entries $\mathcal M_{n,m}=\langle B_{m}, U_{t_{n}}^{\dagger}MU_{t_{n}}\rangle$. Clearly, if $\mathcal M$ is invertible the Bloch vector can be inferred through $\bold{x}_{\rho}=\mathcal M^{-1}\bold{y}$, which is referred to as \emph{information completeness}.

\subsection{Singular controls and information completeness}\label{sec:scandic}
In order to establish a relation between singular controls and information completeness, we first consider the case in which the measured observable is given by the control Hamiltonian, i.e., $M=H_{c}$, which can be interpreted as measuring the response of the control field. By picking the same set of time points $\mathcal T$ as in the tomography case, the singular control  condition \eqref{eq:singularcritical} can be expressed in the operator basis $\{B_{m}\}$ as $\mathcal M \bold{x}_{v}=0$ where $\bold{x}_{v}=(\langle v,B_{1}\rangle,\cdots,\langle v,B_{d^{2}-1}\rangle)$. Thus, if a control field is singular, the corresponding measurement data cannot be informationally complete, as then there does not exist a set of time points for which $\mathcal M$ is invertible. Conversely, if control fields provide information completeness, they cannot be singular controls. We note that while at this point we have not made any assumptions about controllability or control field constraints, generic state reconstruction is only possible if the dynamical Lie algebra is full \cite{d2003quantum}.  

In order to get a more complete picture (i.e., not just a set of points on the landscape), we proceed by assuming that the system is fully controllable, such that at a time $T_{*}$ a random field creates a Haar random evolution. Note that the Haar random time $T_{*}$ can be estimated by mapping the expected dynamics to a Lindbladian dynamics and determining its gap \cite{banchi2017driven}. Consider now a randomly applied field of length $T=(d^{2}-1)T_{*}$ so that at  times $\mathcal T_{*}=\{T_{*},2T_{*},\cdots,(d^{2}-1)T^{*}\}$ a set of $d^{2}-1$ uncorrelated Haar random unitary evolutions $U_{nT^{*}}$ with $n=1,\cdots d^{2}-1$ are created. Then, picking $t_{n}\in \mathcal T_{*}$ yields row vectors $a_{n}=(\langle B_{1},U_{nT^{*}}^{\dagger}H_{c}U_{nT^{*}}\rangle,\cdots,\langle B_{d^{2}-1},U_{nT^{*}}^{\dagger}H_{c}U_{nT^{*}}\rangle)$ of $\mathcal M$ that are formed by statistically independent Hermitian operators $U_{nT^{*}}^{\dagger}H_{c}U_{nT^{*}}$ distributed uniformly within the space of Hermitian operators with the same spectrum as $H_{c}$. According to a standard result from measure theory, the set of row vectors $\{a_{n}\}$ created in this way are linearly independent with probability one. Consequently, the matrix $\mathcal M$ is for almost all control fields of length $T=(d^{2}-1)T_{*}$ invertible. Thus, as long as the control fields allow for creating Haar random evolutions, the set of singular controls form a set of measure zero. We remark here that while the determinant of $\mathcal M$ is an analytical function of the control field parameters, and the zeros form a set of measure zero, the zeros are not necessarily isolated. 
  
 Turning to tomography of a generic state, it follows that expectation value measurements of $H_{c}$ at times $\mathcal T_{*}$ yield for almost all control fields an informationally complete measurement record.  Unitary invariance of the Haar measure then immediately implies that information completeness generically holds when observables that are unitarily conjugate to $H_{c}$, i.e., $M=V^{\dagger}H_{c}V$ with $V\in\text{SU}(d)$, are considered \cite{yang2020complete}. Both results together can be leveraged to find control fields in a learning fashion when system access is limited, for instance, to a single qubit only \cite{lloyd2004universal}.

% \emph{Learning control with a random field} -- 

\subsection{Learning control with a random field}\label{sec:learningcontrol}
 As mentioned in the introduction, optimal controls are typically found in an iterative fashion. Employing for example a gradient type algorithm, in each iteration step $i$ the control field is updated into the direction of the gradient $\frac{\delta J}{\delta f(t)}$ \cite{khaneja2005optimal}. Assuming the control field is piecewise constant on $N$ intervals $\Delta t=T/N$ where its corresponding amplitudes $f(j\Delta t)$ with $j=1,\cdots,N$ are collected in the vector $\bold{f}$, the update step reads $\bold{f}^{(i+1)}=\bold{f}^{(i)}+\alpha \nabla_{\bold{f}}J(\bold{f}^{(i)})$. Here $\nabla_{\bold{J}}$  denotes the gradient with respect to the piecewise constant field amplitudes and $\alpha$ is the step size determining the speed of convergence. In general, there are two routes for carrying out such an iterative search for the optimal controls. The first approach (i) relies on an accurate model describing the controlled system. Here the cost functional and the gradient are \emph{numerically} calculated \cite{machnes2011comparing, qutip}. The second approach (ii), known as learning control \cite{judson1992teaching, li2017hybrid, bukov2018reinforcement, niu2019universal}, is an experimental procedure that does not rest on a model. Instead, at each iteration step the cost functional and the gradient, if needed, are determined through measurements. When system parameters are uncertain, for example when the experimental setting drifts, the quality of the optimal controls obtained through (i) can drop when applied to an experimental implementation. Approach (ii) on the other hand typically requires access to each system component to measure a complete set of observables in order to infer $J$ and $\nabla_{\bold{f}}J$.  
 \begin{figure}[!t]
	\includegraphics[width=0.9\linewidth]{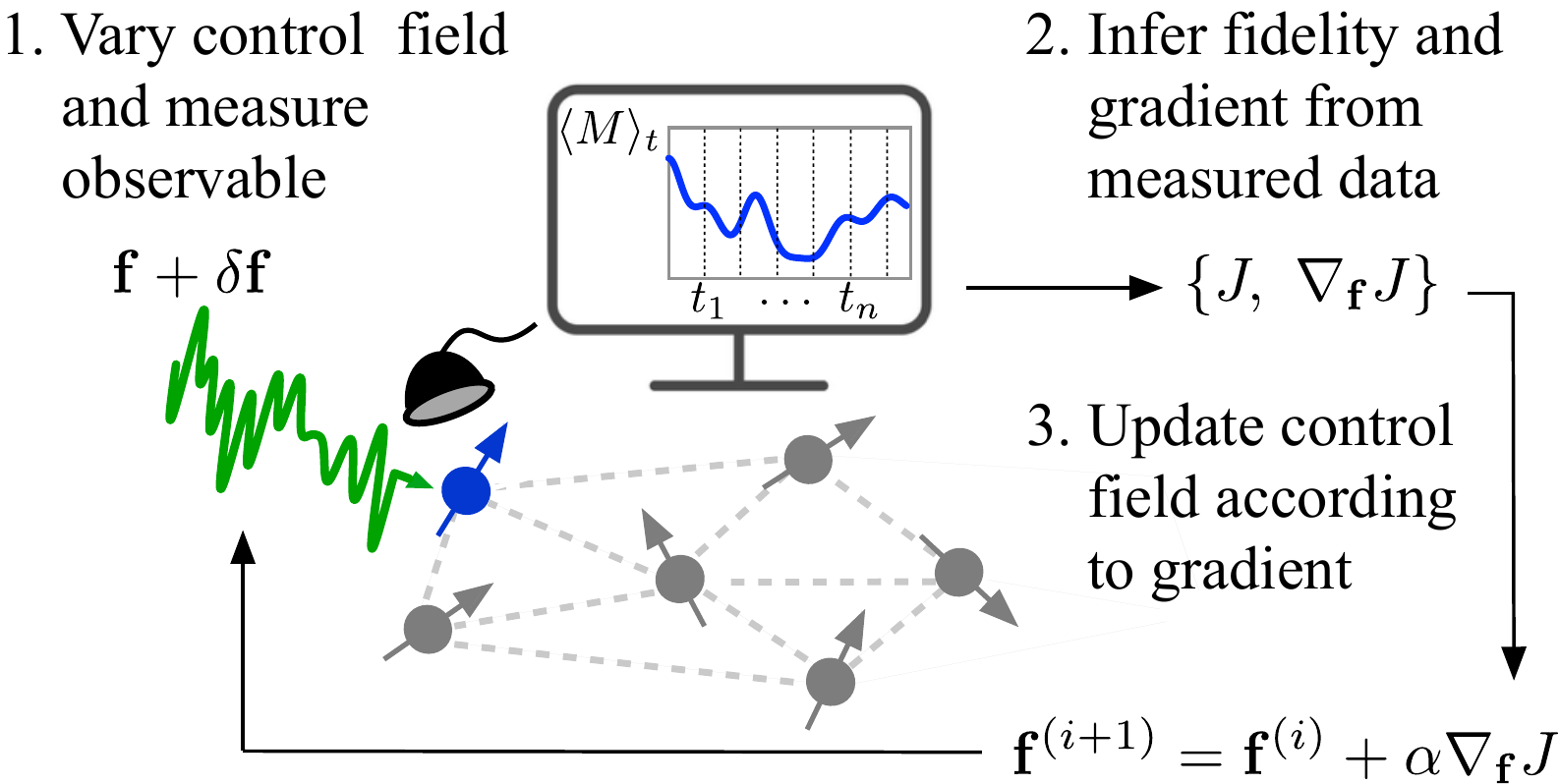}
	\caption{\label{fig3} Schematic representation of how quantum control and quantum tomography can be drawn together to find optimal control fields for preparing a target state when system access is limited to a single qubit (blue). In each  iteration step $i$ a new field $\bold{f}^{(i+1)}$ is determined by adding a random field to $\bold{f}^{(i)}$ to  infer $J$ and $\nabla_{\bold{f}}J$ through expectation value measurements of an observable $M$ at time points $\{t_{1},\cdots,t_{d^{2}-1}\}$. This procedure allows for reconstructing the state of the system created by $\bold{f}^{(i)}$, and thereby $J$, by directly inverting $\mathcal M$ or solving a least squares problem \cite{yang2020complete}. If needed, varying the applied field by $\delta \bold{f}$ then allows for evaluating the gradient $\nabla_{\bold{f}}J$ in a similar fashion, e.g. though finite differences.}  
\end{figure}
 
One way to overcome these challenges is by utilizing a random field. As schematically represented in Figure 1, consider the control problem of preparing a qubit system in a target state $|\psi_{g}\rangle$ and assume that the system is fully controllable. The cost functional is given by the fidelity $J_{\psi_{g}}(\bold{f})=|\langle \psi_{g}|\psi(\bold{f})\rangle|^{2}$ where $|\psi(\bold{f})\rangle=U_{T}(\bold{f})|\phi\rangle$ is the state created from the initial state $|\phi\rangle$ through the solution $U_{T}(\bold{f})$ of the Schr\"odinger equation up to $T$ depending on $\bold{f}$. Picking an initial guess pulse $\bold{f}^{(0)}$, followed by a random field and measuring the time trace of a single qubit observable allows for reconstructing the state $U(\bold{f}^{(0)})|\phi\rangle$ created by the initial guess and thereby the fidelity $J_{\psi_{g}}(\bold{f}^{(0)})$. In the same way the gradient can be estimated, for example by perturbing the control field and using finite differences. Note that this procedure does not require full knowledge of the underlying model. In fact, while knowledge of the system evolution $U_{t}$ is required to form $\mathcal M$, this can be accomplished by also creating a set of random initial states, which allows for inferring the evolution up to some gauge  \cite{blume2013robust, burgarth2012quantum,  blume2017demonstration}. Even when a model is assumed (i.e., in principle one can use approach (i) directly), hybrid approaches  \cite{egger2014adaptive, li2017hybrid, ding2019collaborative, chen2020combining} involving measuring the fidelity and perhaps the gradient have the advantages that faster convergence can be achieved by adjusting $\alpha$ in each iteration step and that moderate uncertainties in the underlying model can be accounted for.

%  \emph{Measurement errors and flat landscapes} -- 
  
  \section{Measurement errors and flat landscapes}\label{sec:errorsandlandscape}
   The practical utility of methods based on learning control relies on having at most moderate noise levels in the measurement record. Not only must the noise be kept at bay, but varying the control field must also allow for sufficiently large variations of the system dynamics, as one must be able to distinguish such changes from the noise. Below we restrict ourself to errors collected in a vector $\bold{\epsilon}$ that disturb the measurement record $\bold{y}$ in a linear fashion i.e., the experimentally observed measurement record takes the form $\tilde{\bold{y}}=\bold{y}+\bold{\epsilon}$. Such errors could for example stem from a finite sample statistics of the expectation values $\langle M\rangle_{t_{n}}$. We remark that thereby we exclude errors that perturb the measurement record in a non-linear fashion, such as noisy control fields or model imperfections. Such errors can be considered analogous to the state preparation and measurement errors known as (SPAM), as pulse and model imperfections perturb the effectively created observables $U_{t_{n}}^{\dagger}MU_{t_{n}}$ used to reconstruct the state. Such SPAM errors motivated the development of self-consistent tomography schemes, e.g., gate set tomography \cite{blume2013robust, blume2017demonstration}, whose ideas can be combined with random-field tomography to reconstruct quantum states without prior knowledge of the Hamiltonian or the random pulse shape, as mentioned earlier in Sec. \ref{sec:learningcontrol}. The analysis and mitigation of non-linear errors as well as the development of  robust schemes goes beyond the scope of this work and will be studied elsewhere.

The difference between the experimentally reconstructed Bloch vector $\tilde{\bold{x}}_{\rho}=\mathcal M^{-1}\bold{\tilde{y}}$ and the true Bloch vector $\bold{x}_{\rho}$ is given by $\Vert \tilde{\bold{x}}_{\rho}-\bold{x}_{\rho}\Vert_{2}=\Vert \mathcal M^{-1}\bold{\epsilon}\Vert_{2}$, where $\Vert\cdot\Vert_{2}$ denotes the $l_{2}$ norm. As such, the ability to reconstruct quantum states in the presence of measurement errors depends on how well $\mathcal M$ is conditioned. Based on an upper bound on the smallest singular value $s_{n}$ of random matrices \cite{rudelson2008least, tatarko2018upper}, in Appendix \ref{sec:app1} we show that for $\mathcal M$ with row vectors uniformly distributed according to the Haar measure there exists a constant $L>0$ such that $\Vert \mathcal M^{-1}\Vert=1/s_{n}(\mathcal M)$ is lower bounded by 
\begin{align}
\label{eq:lowerbound}
\frac{(d^{2}-1)^{3/2}}{L\,|\text{Tr}\{M^{2}\}|}\leq \Vert \mathcal M^{-1}\Vert .  
\end{align}   
This bound suggests that state reconstruction through randomly creating observables, either via a randomly applied control field \cite{yang2020complete} or via gate sequences implementing t-designs \cite{gross2010quantum, ohliger2013efficient}, becomes more sensitive to measurement errors when the system size $d$ increases. This is a consequence of the concentration of measure phenomenon \cite{ledoux2001concentration}, which is encapsulated by the statement that random quantities, such as the randomly created observables, get more and more centered around the mean in high dimensions.  

Another consequence of the concentration of measure phenomenon is the appearance of so-called Barren plateaus \cite{mcclean2018barren} in the control landscape associated to variational algorithms, where the dynamic gradient of the cost functional $J$ can become exponentially small in high dimensions. Using a form of Levy's Lemma \cite{puchala2016distinguishability, ledoux2001concentration}, with details found in Appendix \ref{sec:app2}, we show that for uniformly random target (or initial) states $|\psi_{g}\rangle$ the probability $\mathbb P$ that the gradient $\frac{\delta J_{\psi_{g}}}{\delta f(t)}$ is larger then any value $\kappa$, is upper bounded by     
\begin{align}
\label{eq:bound}
\mathbb P\left[\ \left| \frac{\delta J_{\psi_{g}}}{\delta f(t)}\right|>\kappa \right]\leq 2 \exp\left( \frac{-\kappa^{2} d}{81\pi^{3} E_{\text{max}}^{2}}\right),  
\end{align} 
where $E_\text{max}=\max_{i}\{|E_{i}^{(c)}|\}$ with $\{E_{i}^{(c)}\}$ being the eigenvalues of the control Hamiltonian $H_{c}$. Thus, for most target states the control landscape becomes more and more flat as the system becomes larger. Considering $n$ qubit systems (i.e., $d=2^{n}$) and picking $\kappa=2^{-n/4}$, we see that the probability that the gradient is not exponentially small in the number of qubits $n$ vanishes double exponentially in $n$. We remark that this observation is cost functional-dependent, and consequently, the identification of cost functionals that can mitigate this behavior remains an open challenge \cite{cerezo2020cost}.

Results \eqref{eq:lowerbound} and \eqref{eq:bound} independently emphasize the challenges associated with quantum state reconstruction through randomized schemes and finding optimal controls in high dimensions, respectively. Together, they influence the ability to control complex quantum systems in a learning (or hybrid) fashion. Namely, given that that the probability that a change in the control field causes a change in the cost functional that is larger than $\kappa$ becomes exponentially small, the noise level must be smaller than $\kappa$ to detect such a change. Moreover, if one wants to detect the change using randomly created observables, the bound \eqref{eq:lowerbound} suggests that errors in the corresponding expectation measurements must also be sufficiently small, which again scales with the dimension of the system.

% \emph{Conclusions} -- 
 
 \section{Conclusions}
 In this work, we have identified properties of the control landscape as a common foundation for controlling and reading out quantum systems. In particular, we have shown that the existence of singular controls manifests as an incomplete measurement record of a single observable, as the unitary evolution operator cannot be steered in all directions at these points on the control landscape. Conversely, if a set of control field shapes yields information completeness, they cannot be singular controls. We have shown that if control fields allow for creating Haar random unitary evolutions, then the set of singular controls forms a set of measure zero and consequently, information completeness is a generic property. Building on these results, we have presented a potentially practical experimental procedure, where optimal controls are identified using measurement data obtained from accessing only part of the system.

We have also pointed out potential challenges that may arise as the system dimension is scaled up. Namely, due to the concentration of measure phenomenon, expectations of randomly created observables become more centered around the mean, suggesting that state reconstruction based on random schemes becomes more sensitive to measurement errors as the system size increases. Rooted in the same principle, we further showed that for a cost functional commonly employed for state preparation, most target states yield an exponentially flat landscape. Whether these observations become critical challenges for a given problem is application- and platform-dependent, and thus, as technology progresses, the ability to scale up the system size can be expected to improve.

\emph{Acknowledgements} -- We would like to thank R. B. Wu and A. Magann for insightful comments. C.A.  acknowledges support from the ARO (Grant No. W911NF-19- 1-0382). H.S. acknowledges support from the NSF (Grant No. CHE-1763198).

\newpage
\bibliography{references.bib}

\appendix

\section{Derivation of the bound (4)}\label{sec:app1}
From \cite{rudelson2008least, tatarko2018upper} we have that for a $d\times d$ matrix $A$ with independent and identically distributed random entries $a_{i,j}$ satisfying $\mathbb E[a_{i,j}]=0$ and $\sigma=\mathbb E[a_{i,j}^{2}]=1$ (i.e., mean zero and unit variance), there exist a constant $L$ such that $s_{n}(A)\leq \frac{L}{\sqrt{d}}$ where $s_{n}=1/\Vert A^{-1}\Vert$, with $\Vert \cdot\Vert$ being the operator norm from $l_{2}$ to $l_{2}$, denotes the smallest singular value. 

 In order to use this result we first calculate the expectation and the variance of $\mathcal M_{n,m}=\langle B_{m}, U_{n}^{\dagger}MU_{n}\rangle$ with $U_{n}$ being a Haar random unitary matrix, as well as show that the entries of $\mathcal M$ are independent (i.e., not correlated). To calculate the first and second moments we use well known results \cite{puchala2017symbolic} for the integration with respect to the Haar measure over the unitary group $\text{U}(d)$. In particular, we use   
 
 \begin{align}
 \label{eq:firstm}
 &\int_{\text{U}(d)}U_{i,j}\bar{U}_{i^{\prime},j^{\prime}}dU=\frac{\delta_{i,i^{\prime}}\delta_{j,j^{\prime}}}{d},\\
  \label{eq:secondm}
&\int_{\text{U}(d)}U_{i_{1},j_{1}}U_{i_{2},j_{2}}\bar{U}_{i_{1}^{\prime},j_{1}^{\prime}}\bar{U}_{i_{2}^{\prime},j_{2}^{\prime}}dU \nonumber \\ 
&=\frac{\delta_{i_{1},i_{1}^{\prime}}\delta_{i_{2},i_{2}^{\prime}}\delta_{j_{1},j_{1}^{\prime}}\delta_{j_{2},j_{2}^{\prime}}+\delta_{i_{1},i_{2}^{\prime}}\delta_{i_{2},i_{1}^{\prime}}\delta_{j_{1},j_{2}^{\prime}}\delta_{j_{2},j_{1}^{\prime}}}{d^{2}-1}\nonumber\\ &+\frac{\delta_{i_{1},i_{1}^{\prime}}\delta_{i_{2},i_{2}^{\prime}}\delta_{j_{1},j_{2}^{\prime}}\delta_{j_{2},j_{1}^{\prime}}+\delta_{i_{1},i_{2}^{\prime}}\delta_{i_{2},i_{1}^{\prime}}\delta_{j_{1},j_{1}^{\prime}}\delta_{j_{2},j_{2}^{\prime}}}{d(d^{2}-1)},
 \end{align} 
 where $U_{i,j}$ are the matrix elements of the unitary $U$ and $\bar{U}$ denotes the complex conjugate.
 
 \noindent 
Using \eqref{eq:firstm} the expectation can be calculated and we find  
 \begin{align}
 \mathbb E_{U_{n}}[\mathcal M_{n,m}]&= \mathbb E_{U_{n}}[\text{Tr}\{U_{n}^{\dagger}MU_{n}B_{m}\}]\nonumber \\
 &=\frac{\text{Tr}\{M\}\text{Tr}\{B_{m}\}}{d}=0. \nonumber
 \end{align}
With \eqref{eq:bound} we have that the correlations and the variance are given by 
\begin{align}
&\mathbb E_{U_{n},U_{n^{\prime}}}[\mathcal M_{n,m}\mathcal M_{n^{\prime},m^{\prime}}]\nonumber\\
&=\begin{cases}
\mathbb E_{U_{n}}[\mathcal M_{n,m}]\mathbb E_{U_{n^{\prime}}}[\mathcal M_{n^{\prime},m^{\prime}}]=0,~~~~&\text{for}~n\neq n^{\prime},  \\ 
 \frac{\text{Tr}\{M^{2}\}\delta_{m,m^{\prime}}}{d^{2}-1}, ~~~~&\text{for}~n= n^{\prime}. 
\end{cases} 
 \end{align}
As such, the matrix elements of $\mathcal M$ are uncorrelated and their variance is given by 
\begin{align}
\sigma=\frac{\text{Tr}\{M^{2}\}}{d^{2}-1}.
\end{align}
Defining $\tilde{\mathcal M}=\frac{1}{\sigma}\mathcal M$ allows us to use the above bound to establish $\frac{\sqrt{d^{2}-1}}{L}\leq \Vert \tilde{\mathcal M}^{-1}\Vert $, such that we arrive at our desired result 
\begin{align}
\frac{(d^{2}-1)^{3/2}}{L\,|\text{Tr}\{M^{2}\}|}\leq \Vert \mathcal M^{-1}\Vert. 
\end{align}

\section{Derivation of the bound (5)}\label{sec:app2}
Here we outline how to establish the bound \eqref{eq:bound} using Levy's Lemma given in \cite{puchala2016distinguishability}:\\

\noindent 
\textbf{Levy's Lemma:} \textit{Let $h:S^{N-1}\rightarrow \mathbb R$ be a real valued function on the $N-1$ dimensional Euclidian sphere with Lipschitz constant given by $\lambda=\sup_{x1,x_{2}}\frac{|h(x_{1})-h(x_{2})|}{\Vert x_{1}-x_{2}\Vert_{2}}$. Then, for a uniform random point $x\in S^{N-1}$ and all $\kappa$,}
\begin{align}
\mathbb P[|h(x)-\mathbb E[h]|>\kappa]\leq 2\exp\left(\frac{- N \kappa^{2}}{9 \pi^{2}\lambda^{2}}  \right) \nonumber.
\end{align} 
Levy's Lemma states that the probability to find $h(x)$ with $x$ drawn uniformly random from a $N-1$ dimensional sphere more than $\kappa$ away from the mean $\mathbb E[h]$ is exponentially small. To apply Levy's Lemma we first identify $x$ as the Haar random state $|\psi_{g}\rangle$ on the $S^{2d-1}$ Euclidian sphere and $h(x)$ by the gradient 

\begin{align}
\frac{\delta J_{\psi_{g}}}{\delta f(t)}&=2 \Im[\langle \psi_{g}|U_{T}U_{t}^{\dagger}H_{c}U_{t}|\psi_{0}\rangle\langle\psi_{0}|U_{T}^{\dagger}|\psi_{g}\rangle]. 
\end{align}

In order to calculate the expectation $\mathbb E_{\psi_{g}}\left[\frac{\delta J_{\psi_{g}}}{\delta f(t)}\right]$ and upper bound the Lipschitz constant $\lambda$ we define $h(\psi_{g}):=2 \Im[\langle \psi_{g}|A|\psi_{g}\rangle]$ where $A=U_{T}U_{t}^{\dagger}H_{c}U_{t}|\psi_{0}\rangle\langle\psi_{0}|U_{T}^{\dagger}$. Using \eqref{eq:firstm} from above the expectation can be calculated to be 
\begin{align}
\mathbb E_{\psi_{g}}[h(\psi_{g})]&=-i\mathbb E_{\psi_{g}}[\langle \psi_{g}|A|\psi_{g}\rangle-\overline{\langle \psi_{g}|A|\psi_{g}\rangle}]\nonumber, \\
&=-i(\mathbb E_{\psi_{g}}[\langle \psi_{g}|A|\psi_{g}\rangle]-\overline{\mathbb E_{\psi_{g}}[\langle \psi_{g}|A|\psi_{g}\rangle]})\nonumber, \\
&=-i\left(\frac{\text{Tr}\{A\}}{d}-\frac{\overline{\text{Tr}\{A\}}}{d}\right)\nonumber,  \\
&=0 \nonumber,
\end{align}
where we used in the last step that $\text{Tr}\{A\}=\langle \psi_{0}|U_{t}^{\dagger}H_{c}U_{t}|\psi_{0}\rangle\in \mathbb R$.  What is left is to upper bound the Lipschitz constant $\lambda$. Using 
\begin{align}
|h(\psi_{g}^{(1)})-h(\psi_{g}^{(2)})|&\leq 2|\langle\psi_{g}^{(1)}|A|\psi_{g}^{(1)}\rangle-\langle\psi_{g}^{(2)}|A|\psi_{g}^{(2)}\rangle| \nonumber \\
&\leq 4\Vert A\Vert_{\text{FS}}\Vert |\psi_{g}^{(1)}\rangle-|\psi_{g}^{(2)}\rangle\Vert_{2}\nonumber ,
\end{align}
where $\Vert A\Vert_{\text{FS}}=\sqrt{\text{Tr}\{A^{\dagger}A\}}=\sqrt{\text{Tr}\{|\psi_{0}\rangle\langle\psi_{0}|U_{t}^{\dagger}H_{c}U_{t}\}}\leq E_{\text{max}}$ with $E_{\text{max}}=\max_{i} \{|E_{i}^{(c)}|\}$ and $\{E_{i}^{(c)}\}$ being the eigenvalues of $H_{c}$ we have 
\begin{align}
|h(\psi_{g}^{(1)})-h(\psi_{g}^{(2)})|\leq 4 E_{\text{max}}\Vert|\psi_{g}^{(1)}\rangle-|\psi_{g}^{(2)}\rangle\Vert_{2}\nonumber ,
\end{align}
such that the Lipschitz constant is upper bounded by $\lambda \leq 4 E_{\text{max}}$. We are now able to apply Levy's Lemma to arrive at the desired result \eqref{eq:bound} given in the main body of the manuscript.

\end{document}